\input{aipcheck}
\documentclass[final, numberedheadings]{aipproc}
\usepackage{latexsym}
\usepackage{aas_macros}

\newcommand{\object}[1]{#1}

\newcommand{\be}{ \begin {equation}}
\newcommand{\ee}{ \end {equation}}
\newcommand{\text}[1]{\mbox{#1}}
\bibpunct{(}{)}{;}{a}{,}{,}

\layoutstyle{6x9}

\begin{document}

\title{Formation of multi-planetary systems in turbulent disks}


\author{Hanno Rein}{
  address={University of Cambridge,
  Department of Applied Mathematics and Theoretical Physics,
  Wilberforce Road,
  Cambridge CB3 0WA, United Kingdom}
}

\author{John C. B. Papaloizou}{
}

\begin{abstract}

We summarize the analytic model and numerical simulations of stochastically forced planets in a turbulent disk presented in a recent paper by Rein \& Papaloizou \citep{Rein2008}.  
We identify two modes of libration in systems with planets in mean motion resonance which react differently to random forces.
The \textit{slow mode}, which mostly corresponds to
motion of the angle between the apsidal lines of the two planets, is converted to circulation
more readily than the \textit{fast mode} which is associated with oscillations of the semi-major axes.  

We therefore conclude that stochastic forcing due  to disk turbulence may have played an important role in shaping the configuration of observed systems in mean motion resonance.
For example, it naturally provides a mechanism  for accounting for the HD128311 system 
for which the fast mode librates and the slow mode does not. 

\end{abstract}

\maketitle

\section{Introduction}
Of the recently discovered 336 extrasolar planets, at least 75 are in multiple planet systems \citep{exoplanet}. More than  10\%  of these are in or very close to a resonant configuration where two planets show a mean motion commensurability, with at least five systems in or near a 2:1 resonance \citep{Udry07}.

Resonant configurations can be established by dissipative forces acting on the planets which lead to convergent migration \citep{LeePeale2002}. An anomalous  effective kinematic viscosity~$\nu$, but with considerable uncertainty, has been  inferred from observations of accretion rates onto the central protostars. The magneto rotational instability (MRI) is thought to be responsible for this anomalous value of $\nu$ but the level of MRI turbulence and associated density fluctuations are uncertain. The density fluctuations give rise to random forces on close-by protoplanets.

The influence of stochastic forces on migrating planets was explored first by \cite{NelsonPapaloizou04}.
They considered MRI simulations directly and therefore the simulation ran
only for a relatively small number of orbits.
Recently \cite[and see references therein]{Rein2008} presented a parameterized analytic model which can be used to predict the lifetime of resonant systems as a function of the stochastic diffusion coefficient. We shortly summarize this model in section \ref{sec:sum} and focus on the stability of the GJ876 system in section \ref{sec:gj} and the formation of the HD128311 system in section \ref{sec:form}. We refer the interested reader to \cite{Rein2008} for a more detailed discussion.

\section{Analytic model}\label{sec:sum}

Our goal is to estimate the growth rate of orbital parameters and the lifetime of resonant systems analytically without making any restrictive assumptions about the orbits or the nature of the stochastic forces.
We therefore add an additional term to the full Hamiltonian of the system that is linear in a stochastic force $\mathbf F$. 
The new equations of motion can be calculated by elementary means and expressed in terms of the orbital parameters such as semi-major axis $a$, eccentricity $e$, periastron $\varpi$, mean motion $n$, mean longitude $\lambda$. 
Thus, the additional terms due to the stochastic forces are linear in $\mathbf F$.

We assume that each component of the force $\mathbf F$ on each planet, in cylindrical coordinates, 
satisfies the relation $F_i(t)F_i(t')= \left< F_i^2 \right> g(|t-t'|)  $
 where the autocorrelation function $g(x)$ is such that
 $\int^{\infty}_0 g(x) dx = \tau_c,$
with $\tau_c$ 
being the correlation time and ${\left< F_i^2 \right>}$ the mean square
value of the $i$ component. 
The stochastic forces make quantities they act on undergo a random walk. 
The square of the change of such a quantity $A$ occuring
after a time interval $t$ is given by
$ (\Delta A)^2 = D_i t$, where $D_i = 2\left< F_i^2 \right>\tau_c$ is the diffusion coefficient. 

Considering stochastic forcing on a single isolated planet when $D_i\equiv D$ is independent of $i$, we obtain precise statistical estimates for the 
growth of the orbital parameters 
\begin{eqnarray}
 (\Delta a)^2 = 4 \frac {  D t }{ n^2 }\label{eq:growtha} \quad\quad\quad
 (\Delta e )^{2} = 2.5 \displaystyle\frac{\gamma D t }{ n^2 a^2 }  \label{eq:growthe}\quad\quad\quad
 (\Delta \varpi)^2 = \frac{2.5}{e^2} \displaystyle\frac{\gamma Dt}{n^2 a^2} \label{eq:growthw}.
\end{eqnarray}
When  two planets are involved there are 
two modes of oscillation (libration) which we separate and describe as
fast and slow modes with frequencies $\omega_{lf}$ and $\omega_{ls}$.  
To summarize our results, the ultimate lifetime of a resonant configuration of planets is determined
by the time taken for the fast angle to achieve circulation. We determine this lifetime
using equations similar to those described above but for the libration amplitude, obtaining
\begin{eqnarray}
 t_f&=&2.4\times 10^{-4}\left(\frac {a_1^2n_1^4}{\langle F_i^2\rangle}\right)
 \left(\frac{1}{8n_1\tau_c}\right) 
  \left( \frac{8.5\omega_{lf}\sqrt{q_{GJ}}}{n_1\sqrt{q}}\right)^2\frac{q}{q_{GJ}} P_1.\label{RESEQ}
\end{eqnarray}
Here the first quantity in brackets represents the ratio of the square of the central force 
to the mean square stochastic force acting on the outer planet. The other quantities in brackets
 are expected to be of order unity, while the last factor 
$q/q_{GJ}$ is the ratio of the planet mass to the stellar mass compared to the same quantity in the \object{GJ876} system. $P_1$ is the period, with the subscript $1$ denoting the outer planet. 
We showed that this is in very good agreement with numerical results \citep{Rein2008}.

\section{Stability of GJ876}\label{sec:gj}
\begin{figure}
\centering
\includegraphics[width=0.5\columnwidth]{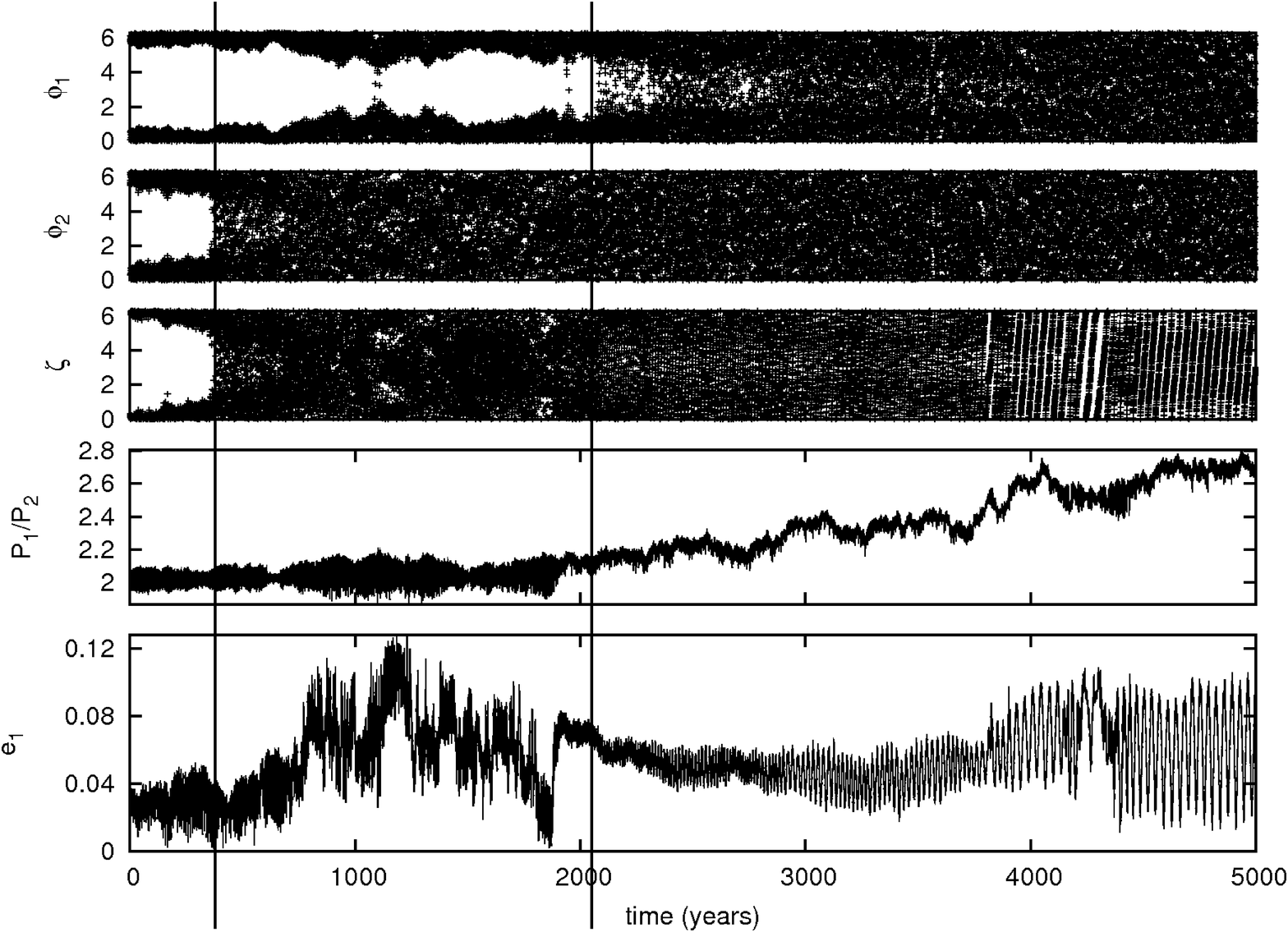}
\includegraphics[width=0.5\columnwidth]{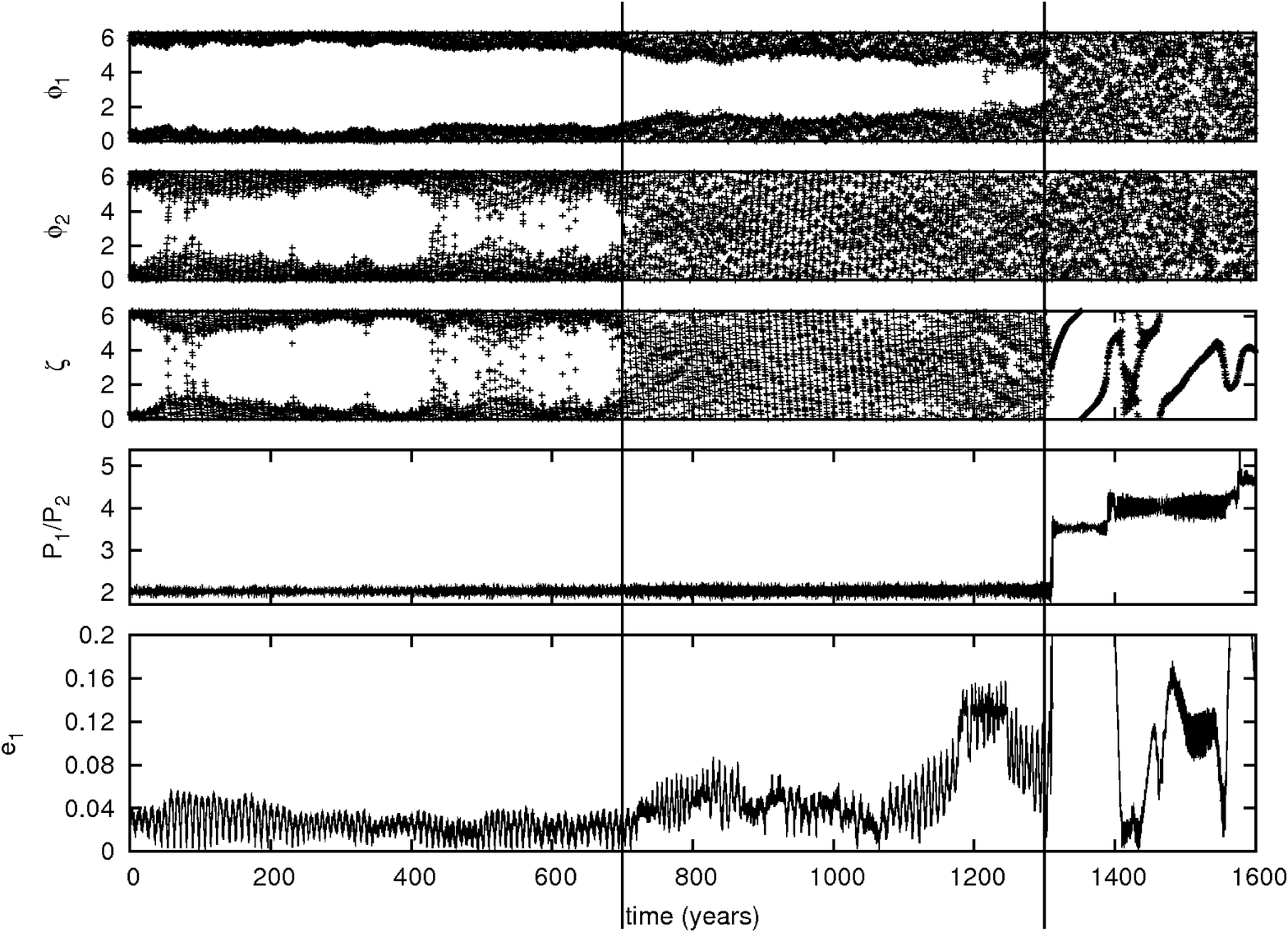}
\caption{Time evolution of the resonant angles $\phi_1=2\lambda_1-\lambda_2-\varpi_2$ (fast mode), $\phi_2=2\lambda_1-\lambda_2-\varpi_1$ and $\zeta=\varpi_2-\varpi_1$ (slow mode), 
the period ratio $P_1/P_2$ and the eccentricity, $e_1,$
 in the  \object{GJ876} system. 
The vertical lines indicate when the angles enter circulation
for a prolonged period.
 The realization illustrated in the right panel
  scatters shortly after the fast libration angle, $\phi_1$ goes
 into circularization. }
 \label{fig:twoplanets_turb_break}
\end{figure}

We have performed numerical simulations of two planet systems that allow for the  incorporation
of additional stochastic forces with the properties described above. 
In order to mimic the effects of 
turbulence it is necessary to calibrate these forces with reference to MHD simulations. 
The only free parameters 
are the mean square value of the force components 
per unit mass  $\left< F_i^2 \right>$ 
and the auto correlation time $\tau_c$. 
 The numerical noise generator that we implemented, uses a discrete first order Markov process \citep{KASDIN} to generate a correlated and continuous force. 
We conclude from numerical MHD simulations that the  natural scale for the diffusion coefficient is
specified through 
\begin{equation}
 D_0 \approx 10^{-5}\left( \frac{r}{1\mbox{ AU}}\right)^{-3/2}
  \left(\frac{M_*}{1 \mbox{ M}_\odot}\right)^{-1/2} \frac{ \mbox{cm}^2}{\mbox{s}^{3}}.\label{eq:scalingd} 
\end{equation}
Of course we emphasize  that the value of this quantity is very uncertain.

In order to illustrate the 
 evolutionary sequence we plot results
  for two realizations of the \object{GJ876} system
in figure \ref{fig:twoplanets_turb_break}.
  For these runs we adopted a diffusion coefficient of
$D = 0.42 \text{cm}^2/\text{s}^{3}.$ 
In this context we note that reducing $D$ increases
the evolutionary time which has been found, both analyticaly and numerically 
 to be~$\propto~1/D$.
The times at which the transition 
from libration to circulation occurs are different for the slow and fast angles and
indicated by vertical lines in figure \ref{fig:twoplanets_turb_break}.
The tendency for the occurrence of very small
values of $e_1$ can be associated with transitions to circulation
of the slow mode, $\zeta$ being the angle between pericenters. 
We have verified this by considering the results from simulations which started with a larger  value of $e_1$. As expected,
the attainment of circulation of the slow angle takes longer
in this case.
Also as expected, the time when $\phi_1$ attains circulation is not affected by the change in $e_1$.

\section{Formation of HD128311}\label{sec:form}
\begin{figure}
\centering
\includegraphics[width=0.5\columnwidth]{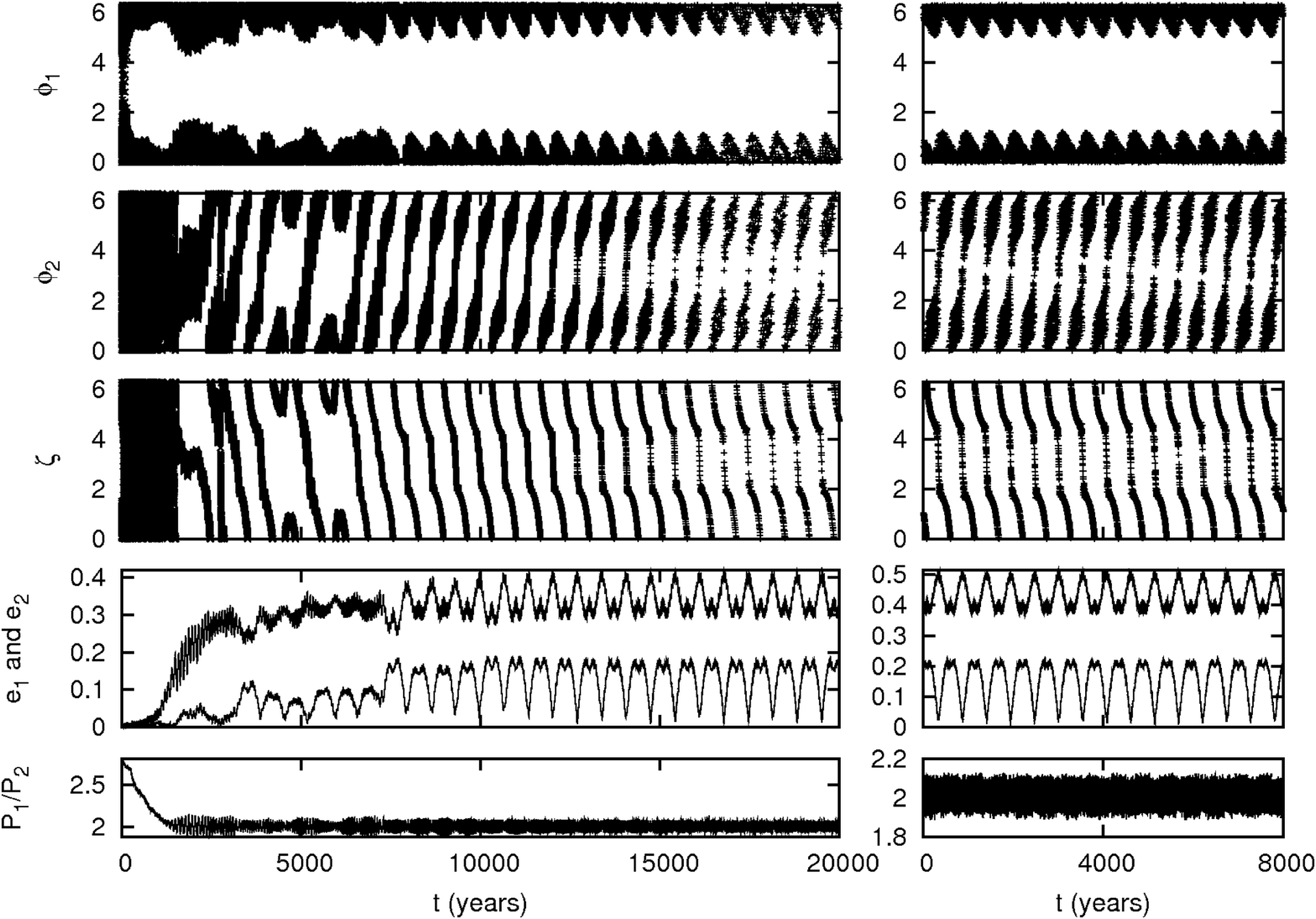}
\includegraphics[width=0.5\columnwidth]{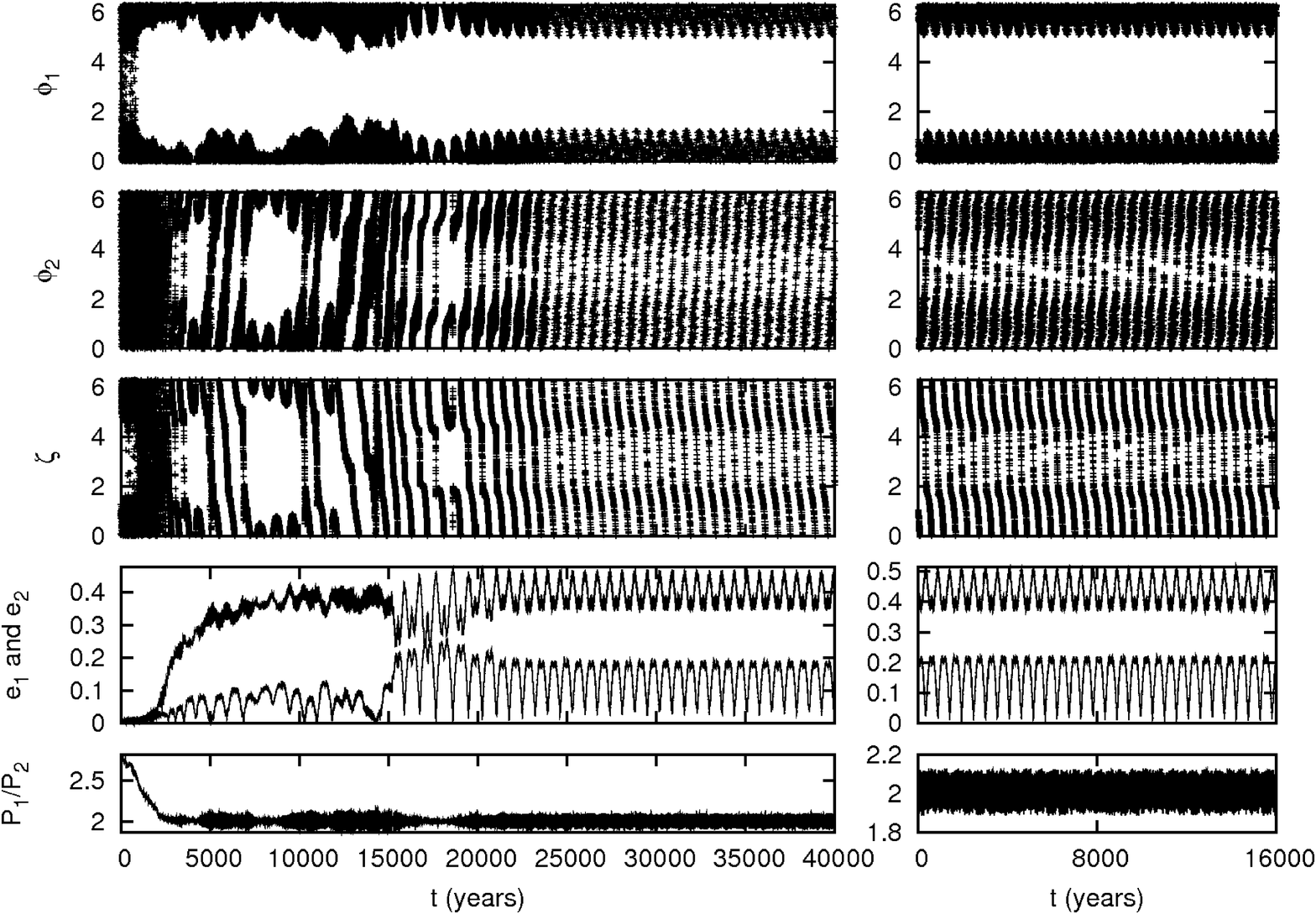}
\caption{
The plots show two different formation scenarios of \object{HD128311} including turbulence and migration. 
We plot the observed system on the right hand side of each plot. 
The plots show the resonant angles $\phi_1$, $\phi_2$ and $\zeta$, the eccentricities $e_1$ and $e_2$ as well as the period ratio $P_1/P_2$. 
 \label{fig:hd128311}
 \label{fig:hd128311_1}
 \label{fig:hd128311_2}
}
\end{figure}

The ideas presented above can help us to understand
the orbital configuration of 
the planetary system \object{HD128311}.    This system is in a 2:1 mean motion resonance with the angle $\phi_1$  librating  and the angle $\zeta$ circulating (no apsidal corotation) \citep{Vogt2005}. 

It can be shown \citep{LeePeale2002} that the planets should exhibit apsidal corotation if the 
commensurability was formed  by convergent migration only.
Accordingly, it has been suggested \citep{SandorKley06} that an additional perturbing event (a close encounter with an additional planet) could produce orbital parameters similar to the observed ones.
This perturbation is needed to alter the behaviour of  $\zeta,$ so that it undergoes circulation rather than libration. 

We showed above that stochastic forcing that results from turbulence driven by the MRI readily produces systems with commensurabilities without apsidal corotation. 
This suggests that a scenario which forms the resonance through disk induced migration might readily produce commensurable systems without apsidal corotation if stochastic forcing is included.
We present such formation scenarios that do not invoke artificial perturbation events in figure \ref{fig:hd128311}. 
We indeed find that model systems with orbital parameters resembling the observed ones are formed without difficulty and we are able to obtain similar final states for a wide range of migration parameters.

With radial velocity measurements continuously improving and dedicated telescopes such as \textit{Kepler} many exoplanets will be discovered within the next few years. 
That will allow us to assess the role of stochastic forcing on protoplanets in a statistical manner.

\bibliographystyle{apalike}

\end{document}